\documentstyle[12pt]{article}
\def\beq{\begin{equation}}
\def\eeq{\end{equation}}
\def\pl{\partial}
\def\na{\nabla}
\def\al{\alpha}

\def\Ga{\Gamma}
\def\de{\delta}
\def\De{\Delta}
\def\si{\sigma}
\def\te{\theta}
\def\La{\Lambda}
\def\lam{\lambda}
\def\Om{\Omega}

\def\ep{\epsilon}

\def\sq{\sqrt}
\def\sqg{\sqrt{G}}

\def\l{\left (}
\def\r{\right )}
\def\fr{\frac}
\def\la{\label}
\def\hs{\hspace}
\def\inf{\infty}
\def\ran{\rangle}
\def\lan{\langle}

\begin{document}

\begin{titlepage}
%\begin{flushright}
%BA-??-??\\
%March 13, 2001 \\
%hep-ph/9811xxx
%\end{flushright}

\begin{center}
{\Large\bf   Axionic Domain Wall and Warped Geometry
}
\end{center}
\vspace{0.5cm}
\begin{center}
{\large Qaisar Shafi$^{a}$\footnote {E-mail address:
shafi@bartol.udel.edu} {}~and
{}~Zurab Tavartkiladze$^{b}$\footnote {E-mail address:
z\_tavart@osgf.ge} }
\vspace{0.5cm}

$^a${\em Bartol Research Institute, University of Delaware,
Newark, DE 19716, USA \\

$^b$ Institute of Physics, Georgian Academy of Sciences,
380077 Tbilisi, Georgia}\\
\end{center}

\vspace{1.0cm}

\begin{abstract}
We discuss how a three-brane with an associated non-factorizable (warped)
geometry can emerge from a five dimensional theory of gravity coupled to
a complex scalar field. The system possesses a discrete $Z_2$ symmetry,
whose spontaneous breaking yields an 'axionic' three-brane and a warped
metric.
Analytic solutions for the wall profile and warp factor are presented.
The Kaluza-Klein decomposition and some related issues are also discussed.

\end{abstract}

\end{titlepage}
%\dspace

\section{Introduction}

Theories with extra spacelike dimensions have recently attracted 
a great deal of attention. It was observed \cite{tevgr1, tevgr2} that for suitably
large extra dimensions, it is possible to lower the fundamental mass scale 
of gravity $M_f$
down to a few TeV. This suggests a new way for a solution of the gauge hierarchy
problem without invoking supersymmetry (SUSY). In this approach all the standard model
particles are localized on a $3$-brane, and only gravity propagates in
the bulk. Assuming that the $n$ compact dimensions have a typical size
$R$, the four dimensional Planck scale is expressed as:

\beq
M_{\rm Pl}^2\sim M_f^{2+n}R^n~.
\la{plrad}
\eeq
In order to reproduce the correct behavior of gravity one should take
$R\stackrel{<}{_{\sim}}$~mm (the behavior of 
gravity at this distance is now being studied). Interestingly, already for $n=2$,
$R\sim 1$~mm and $M_f\sim$~few$\cdot $TeV, $M_{\rm Pl}$ has the required
magnitude $\sim 10^{19}$~GeV. Detailed studies of phenomenological and
astrophysical implications of these models, were presented in \cite{tevgr2}.
We note that our Universe
as a membrane embedded in higher dimensional spacetime was also
considered in earlier works \cite{univ}.

An alternative solution of the gauge hierarchy problem invoking an extra
dimension was presented in \cite{RS}. The desired mass
hierarchy is generated through a non-factorizable metric obtained from 
higher dimensional gravity (see also \cite{gogb}). In
the
minimal setting \cite{RS} there are two three branes - hidden and visible,
separated by an appropriate distance. The non-factorizable 
metric is given by:

\beq
ds^2=e^{-2k|y|}ds^2_{3+1}-dy^2~,
\la{warpmet}
\eeq
where $y$ denotes the fifth spacelike dimension, $ds^2_{3+1}$ is the ordinary
$4$D interval, and $k$ is a mass parameter close to the 
fundamental scale $M_f$. On the visible brane all mass parameters are rescaled
due to the warp factor in (2), such that $m_{\rm vis}=M_fe^{-k|y_0|}$ ($y_0$ is
the distance
between branes). For $M_f\sim 10^{19}$~GeV and $k|y_0|\simeq 37$ one 
finds that 
$m_{\rm vis}\sim $~few$\cdot$~TeV, the desired magnitude. It was
also shown that
Newton's law still holds on the visible brane. It is worth noting that in this
approach the extra dimension can be infinite \cite{open1}, provided
it's volume remains finite.
Generalization of this non-factorizable model to scenarios with open
codimensions and with
intersecting multiple branes was presented in \cite{open2}. 
%For various
%theoretical features and phenomenological implications, of models with
%warped geometry, see papers \cite{?} and also relevant references
%therein.

It is clearly important to inquire about the origin of the 3-branes in the above
scheme with the warped metric.
In this paper we present one such scenario with a complex scalar field coupled to
$5$D gravity. The theory possesses $5$D Poincare invariance and $Z_2$
discrete symmetry. The 3-brane and warped geometry emerge dynamically from
spontaneous breaking of the $Z_2$ symmetry. The 3-brane describes a topologically
stable domain wall, an axion-type
solution of the sine-Gordon equation in curved space-time. Analytical
solutions for the domain wall profile and warp factor are presented.
As expected, the $5$D space turns out to be
Anti-de-Sitter (AdS). Questions of compactification, Kaluza-Klein (KK) 
decomposition,
graviphoton mass and other related issues are also discussed.

 \section{The Model}

In this section we will consider higher dimensional ($D=5$) gravity plus
a complex scalar field which turns out to possess a non-factorizable solution 
of equation (\ref{warpmet}). The motivation
for the
choice of complex scalar is that with the help of $Z_2$ symmetry we naturally
obtain a potential with a cosine profile \cite{singor} familiar from axion models. 
This yields a non
trivial analytical solution for the $\te$-domain wall whose core can be
identified as a $3$-brane.

\subsection{Complex Scalar Coupled to $5$D Gravity}

Consider $5$D gravity coupled to a complex scalar 
field\footnote{For higher dimensional non-factorizable scenarios, extended
with real scalar fields, see \cite{singor}-\cite{pars}.} $\Phi $
through the action

\beq
S=\int d^5x \sqg\l -\fr{1}{2}M^3 R-\La +{\cal L} (\Phi )\r~,
\la{act}
\eeq
with 

\beq
{\cal L}(\Phi )=\fr{1}{2}G^{AB}\l \pl_A\Phi^*\pl_B\Phi+
\pl_B\Phi^*\pl_A\Phi\r -V(\Phi)~.
\la{lag}
\eeq
Here $G_{AB}$ is the $5$D metric tensor and $G={\rm Det}G_{AB}$ 
($A, B=1,\cdots ,5$).
The Einstein equation derived from (\ref{act}) is given by

$$
R_{AB}-\fr{1}{2}G_{AB}R-\fr{\La}{M^3}G_{AB}=\fr{V}{M^3}G_{AB}+
\fr{1}{M^3}(\pl_A \Phi^*\pl_B\Phi+\pl_B \Phi^*\pl_A\Phi)-
$$
\beq
\fr{1}{2M^3}G_{AB}G^{CD}(\pl_C \Phi^*\pl_D\Phi+
\pl_D\Phi^*\pl_C\Phi)~,
\la{eins}
\eeq
while the equation of motion for $\Phi $ follows from

\beq
\fr{\de {\cal L}}{\de \Phi}=
\fr{1}{\sqg}\pl_A\l \sqg \fr{\de {\cal L}}{\de (\pl_A\Phi)}\r~.
\la{eqm}
\eeq
Terms on the right hand
side of (\ref{eins}) effectively play the role of energy-momentum 
tensor $T_{AB}$, which will be the source for the dynamical generation of the
3- brane and yield a non-factorizable geometry.

Before proceeding to the specific model, which fixes $V(\Phi )$, let us derive the appropriate
equations of motion [from (\ref{eins}), (\ref{eqm})]. We are
looking for a metric of the form: 

\beq
G_{AB}={\rm Diag}\l A(y), -A(y), -A(y), -A(y), -1 \r~,
\label{5dtens}
\eeq
which conserves $4$D Poincare invariance:

\beq
ds^2=A(y)\bar{g}_{\mu \nu}dx^{\mu }dx^{\nu }-dy^2~,~~~~~
\bar{g}_{\mu \nu}=\eta_{\mu \nu}+\bar{h}_{\mu \nu}~,
\la{redmet}
\eeq
where

\beq
\eta_{\mu \nu}={\rm Diag}(1, -1, -1, -1)~
\label{4dtens}
\eeq
and $\bar{h}_{\mu \nu}$ denotes the $4$D graviton ($\mu ,\nu =1,\cdots ,4$).
The (1, 1) and (5, 5) components of (\ref{eins}) respectively give

\beq
\fr{A''}{A}=
-\fr{2}{3}\fr{\La+V}{M^3}-\fr{2}{3 M^3}(\Phi^*)'\Phi'~,
\la{ain11}
\eeq

\beq
\l \fr {A'}{A} \r^2=
-\fr{2}{3}\fr{\Lambda+V}{M^3}+\fr{2}{3M^3}(\Phi^*)'\Phi'~,
\la{ain55}
\eeq
where primes denote derivatives with respect to the fifth coordinate $y$.
Subtracting (\ref{ain55}) and (\ref{ain11}), we get:

\beq
-\fr{A''}{A}+\l \fr {A'}{A} \r^2=
\fr{4}{3 M^3}(\Phi^*)'\Phi'~.
\la{ain1}
\eeq
Using the substitutions:

\beq
\Phi=v\cdot e^{{\rm i}~\te }~,
\la{par1}
\eeq

\beq
A=A_0\cdot e^{-\si }~
\la{par2},
\eeq
and assuming that $v$ in (\ref{par1}) does not depend on $y$ [see
discussion in sec. (1.2)], from
(\ref{ain1}) and (\ref{ain55}) we derive:

\beq
\si\hs{0.5mm}'\hs{0.6mm}'=\fr{4 v^2}{3 M^3}~\te\hs{0.5mm}'\hs{0.7mm}^2~,
\la{eq1}
\eeq

\beq
\si\hs{0.5mm}'\hs{0.7mm}^2=-\fr{2}{3}\fr{\La+V}{M^3}+
\fr{2 v^2}{3M^3}~\te\hs{0.6mm}'\hs{0.7mm}^2~.
\la{eq2}
\eeq
With our assumption $v={\rm const}.$, from (\ref{eqm}) we obtain the 
equation of motion for $\te$:

\beq
2v^2~\te\hs{0.5mm}'\hs{0.6mm}'-
4v^2~\si\hs{0.6mm}'~ \te\hs{0.5mm}'-\fr{\pl V}{\pl \te}=0~.
\la{eq3}
\eeq

The three equations (\ref{eq1})-(\ref{eq3}) are not independent.
Namely, differentiating (\ref{eq2}) and using (\ref{eq1}), we obtain
(\ref{eq3}). However, in order for  these three equations to have a solution,
one fine tuning between the parameters is
unavoidable. This can be seen from the following discussion:
Solving equations (\ref{eq2}) and (\ref{eq3}) we have three parameters
of integration $\te (y_0)$, $\te \hs{0.5mm}'(y_0) $ and $\si (y_0)$,
where $y_0$ is some arbitrary point (In principle there is also a fourth parameter which
expresses translation invariance.  But it is irrelevant, since the
equations are invariant under translations). $\si (y_0)$ is also
irrelevant, since the equations contain only derivatives of $\si $.
{}From (\ref{eq1}), $\te (y_0)$ is also irrelevant. Since for the brane
solution
we have to impose the condition $\te \hs{0.5mm}'(\inf )=0$, the
third parameter $\te \hs{0.5mm}' (y_0)$ is fixed from this condition.
Therefore, there remains no free parameters, and for satisfying
(\ref{eq1}), one fine tuning must be done
(for detailed discussions about this issue see 
\cite{pars}).
This will be explicitly seen
for the model discussed below.

\subsection{Axionic Brane and Warped Geometry}

We introduce a $Z_2$ symmetry under which 
$\Phi \to -\Phi$. The relevant potential is given by

\beq
V=\fr{\lam_1}{4}(\Phi^*\Phi-v^2)^2-\fr{\lam }{2}(\Phi^2+\Phi^{*2})~.
\la{pot}
\eeq
The first term in (\ref{pot}) is $(U(1))$ invariant under
$\Phi \to e^{{\rm i}\te }\Phi $, while the last term explicitly breaks it to
$Z_2$. This avoids the appearance of a Goldstone mode because of non-zero
$\Phi $ VEV\footnote{For models with $Z_2$ replacing the PQ
symmetry and avoiding an undesirable axion, see papers \cite{z2, ourz2},
where
various phenomenological and cosmological implications are also studied.}.
We restrict our attention in (\ref{pot}) to terms needed to implement the
scenario. The  couplings $\Phi^4+\Phi^{*4}$ can be included if so desired,
but this makes analytic calculations more difficult. As noted in 
\cite{ourz2}, such
terms are absent in some models. Higher powers in $\Phi $ and 
$\Phi^{*} $ would complicate the discussion even further.
We assume that the U(1) 
violating term is such that

\beq
\lam_1 v^2\gg \lam~.
\la{cond}
\eeq
Therefore, the VEV
$\lan |\Phi |\ran $ is mainly determined by the first term in
(\ref{pot}),

\beq
\lan \Phi^*\Phi \ran \simeq v^2~,
\la{modul}
\eeq
{}from which
\beq
\Phi\simeq v~e^{{\rm i}\hs{0.5mm}\te }~.
\la{aprsol}
\eeq
Substituting (\ref{aprsol}) in  (\ref{pot}), the $\te $ dependent part of
the potential is given by
\beq
V_{\te }=-\lam v^2\cos 2\te ~.
\la{potte}
\eeq
This type of potential was also used for brane formation in
\cite{singor}. In our case we have obtained it through a $Z_2$ symmetry
acting on a complex scalar field $\Phi $.
Assuming $\lam >0$, (\ref{potte}) acquires its minima for
$\te=0, ~\pi $.
The $\lan \te \ran $ VEV breaks the symmetry $\te \to -\te $. This causes
the creation
of topologically stable domain wall. The wall is stretched between
two energetically degenerate minima, $\te =0$ and $\te =\pi $. With
assumption (\ref{cond}) it is consistent to consider $v$ to be (essentially)
$y$-independent.

Introducing the dimensionless coordinate $\xi $

\beq
\xi =\sq{2\lam }~y ~,
\la{cor}
\eeq
(\ref{eq1}) and (\ref{eq3}) respectively become:

\beq
2 \fr {\pl^2 \te}{\pl \xi^2}-
4 \fr {\pl \si}{\pl \xi}\fr{\pl \te}{\pl \xi}-\sin 2\te=0 ~,
\la{singor}
\eeq

\beq
\fr {\pl^2 \si }{\pl \xi^2}=\al \l \fr{\pl \te}{\pl \xi} \r^2~,
\la{eqsi}
\eeq
where
\beq
\al=\fr{4v^2}{3M^3}~.
\la{alp}
\eeq
Nontrivial solutions of (\ref{singor}) and (\ref{eqsi}), with boundary 
conditions

\beq
\te (-\inf )=0~,~~~\te (+\inf )=\pi ~,~~~
\si (\pm \inf )\propto \pm y~,
\la{bouncon}
\eeq 
[Note that due to the breaking of the U(1) symmetry to $Z_2$ in
(\ref{pot}),
the wall here is not 'bounded by strings', a phenomenon encountered in
$SO(10)$ and axion models \cite{kib}.]
will indicate the existence of 'warped' geometry and the 
axion(or ${\te}$)-brane
(since $\lan \te \ran $ breaks $5$D invariance). The point
$\te=\fr{\pi }{2}$ will be identified as the location of the 
$3$-brane describing $4$D theory. 

Using the substitution

\beq
\te=2 \arctan f(\xi )~,
\la{f}
\eeq
(\ref{singor}), (\ref{eqsi}) can be rewritten as
\beq
-(f^2-1)f''+2f(f''f-f'^2)-2\si'(f^2+1)f'+f(f^2-1)=0~,
\la{eqf}
\eeq

\beq
(f^2+1)^2\si''=4\al f'^2~,
\la{eqsif}
\eeq
where primes denote derivatives with $\xi $. The form for $f$

\beq
f=ae^{m\xi}~,
\la{fexp}
\eeq
is a reasonable choice, 
where the parameters $a, m >0$ are undetermined for the time being.
Substituting
(\ref{fexp}) in (\ref{eqsif}), the latter can be integrated:

\beq
\si'=s_0-2\al m\fr{1}{1+f^2}~,
\la{si1}
\eeq
where $s_0$ is some constant. Substituting (\ref{si1}) into (\ref{eqf})
and taking into account that $f''=mf'=m^2f$, we find:

\beq
-(f^2-1)m^2-2m[s_0(f^2+1)-2\al m]+f^2-1=0~.
\la{powf}
\eeq
Comparing appropriate powers of $f$ in (\ref{powf}), it is easy to verify
that (\ref{powf}) is satisfied if

\beq
m=\fr{1}{\sq{1+2\al}}~,~~~~s_0=\fr{\al}{\sq{1+2\al}}~.
\la{ms}
\eeq
Integration of (\ref{si1}) gives

\beq
\si=\alpha \ln[\cosh (m\xi+\delta )]+\ln C~,~~~~~\delta=\ln a
\la{si}
\eeq
($C=$constant and we have taken into account (\ref{ms})).

Finally, for $\te $ and the warp factor $A(=A_0e^{-\si })$ we will
have:

\beq
\te=2\arctan (ae^{m\xi})~,
\la{solte}
\eeq

\beq
A=A_0[\cosh (m\xi +\delta )]^{-\al}~.
\la{solA}
\eeq
where the constant $C$ is now absorbed in $A_0$, 
and $a$ still remains undetermined, which reflects
translational invariance in the fifth direction $\xi $ ($y$).

Let us note here that these solutions are obtained for $\lam>0$
and a negative sign in front of the last term in (\ref{pot}). In case of
a positive sign, the potential is minimized for $\te =\pm \fr{\pi}{2}$,
and instead of the solution (\ref{solte}), we would have 
$\tilde{\te}=\te-\fr{\pi }{2}$. Indeed,  under these modifications, 
equations (\ref{singor}), (\ref{eqsi}) are satisfied [for this case the
sign in front of $\sin \tilde{\te}$ in (\ref{singor}) will be positive,
which reflects a change of sign of the last term in (\ref{pot})].

{}From (\ref{solA}), taking into account (\ref{ms}), we will get
the desirable asymptotic forms for $A$:

$$
A\sim e^{s_0\xi}~,~~~~~~~\xi\to -\infty~,
$$
\beq
A\sim e^{-s_0\xi}~,~~~~~~~\xi\to +\infty~.
\la{asymp}
\eeq
The solutions (\ref{si}), (\ref{solte}) should also satisfy (\ref{eq2}), which
in terms of $\xi $ has the form

 \beq
\si\hs{0.5mm}'\hs{0.5mm}^2=-\fr{\La +V}{3\lam M^3}+
\fr{\al }{2}\te\hs{0.5mm}'\hs{0.5mm}^2~.
\la{eqsi31}
\eeq
{}From (\ref{si1}), (\ref{solte}) and (\ref{ms}) we have

\beq 
\si\hs{0.5mm}'=\al m\fr{f^2-1}{f^2+1}~,~~~
\te\hs{0.5mm}'=\fr{2mf}{f^2+1}~,~~
\cos 2\te=1-\fr{8f^2}{(f^2+1)^2}~.
\la{subs}
\eeq
Substituting all of this in (\ref{eqsi31}), we can see that the latter is
satisfied if
\beq 
\La =\lam v^2(1-4\al m^2)=\lam v^2\fr{1-2\al }{1+2\al}~.
\la{Lam}
\eeq
Therefore, as we previously mentioned, one fine tuning between the
parameters of the theory is necessary.
The effective $5$D cosmological constant is determined to be

\beq 
\La_{eff}=\La+\lan V \ran=\La+V_{\te}(\te=0,~\pi)=
-4\lam v^2\fr{\al }{1+2\al }~.
\la{5Dcos}
\eeq
As expected, the initial $5$D space-time is AdS.

The warp factor (\ref{solA}) reaches its maximum at
$\xi_0=-\ln a/m$ and decays exponentially far from $\xi_0$. For a
realistic model which solves the gauge hierarchy problem, we may regard the axion
wall as a hidden brane, located at $\xi_0$. By placing the
visible brane (which can describe our $4$D Universe) at a distance
$\De \xi \simeq 74/(\al m)$ from $\xi_0$,  all masses on the visible
brane will be rescaled as 
$m_{\rm vis}=M\cdot A(\xi_0+\De \xi)^{1/2}\simeq
M\cdot e^{-\al m\De \xi/2}\sim 10^{-16}\cdot M$. For $M\sim
10^{19}$~GeV, the desired scale $m_{\rm vis}\sim $~few$\cdot$~TeV will
be naturally generated.

\section{Kaluza-Klein Decomposition}

In this section we will present the KK reduction of the $5$D model to $4$D.
We will calculate the graviphoton [($\mu , 5$) component of
metric tensor] mass, as well as the effective $4$D Planck scale and the radius
of the extra
dimension. For KK reduction within models with non-factorizable geometry,
also see papers in \cite{KKdec1, KKdec2}, while for works addressing the
effects of
spontaneous breaking of higher Poincare invariance, Goldstone phenomenon and
other relevant issues within models with non-warped geometry, see
\cite{goldth}.    

For performing a KK decomposition, it is convenient to
rewrite the metric in (\ref{redmet}) in a conformally 'flat' form:

\beq
ds^2=
%\Om^2(z)(\eta_{\mu \nu} dx^{\mu } dx^{\nu }-dz^2)=
\Om^2(z)g_{MN}dx^Mdx^N~,
\la{flmet}
\eeq
where: 

\beq
dz=A^{-1/2}(y) {dy}~,~~~~~~\Om^2(z)=A\l y(z)\r~,
\la{conf1}
\eeq

\beq
G_{MN}=\Om^2g_{MN}~.
\la{conf2}
\eeq
With the standard KK decomposition 

\begin{equation}
\begin{array}{cc}
% & {\begin{array}{cc}
%{\cal N}_1~&\,\,{\cal N}_2~~~~~~
%\end{array}}\\ \vspace{2mm}
%\begin{array}{c}
%\bar 5_1\\ \bar 5_2 \\ \bar 5_3
%\end{array}
g_{MN}=\!\!\!\! & \left (\begin{array}{ccc}
\hs{-1mm}\bar g_{\mu \nu}-k^2 A_{\mu }A_{\nu }~, &
\hs{-0.5mm}k A_{\mu }
\\
\hs{-1mm}k A_{\nu }~, &\hs{-1mm} -1
\end{array}\hs{-1.5mm}\right)\hs{0.1mm},
\end{array} 
\begin{array}{cc}
% &
% {\begin{array}{cc}
%{\cal N}_1&\,\,
%{\cal N}_2~~~
%\end{array}}\\ \vspace{2mm}
%\begin{array}{c}
%{\cal N}_1 \\ {\cal N}_2
%
%\end{array}
g^{MN}=\!\!\!\!\! &{\left(\begin{array}{ccc}
\hs{-1mm}\bar g^{\mu \nu}
 &\,kA^{\mu }
\\
\hs{-1mm} kA^{\nu}
&\,k^2A_{\al }A^{\al }-1
\end{array}\hs{-1.5mm}\right)\hs{0.1mm},
%\left(\frac{X}{M_P}\right)^{r'}
}
\end{array}
\label{tensors}
\end{equation}
where $A_{\mu }$ is the graviphoton, equation (\ref{flmet})
reads

\beq
ds^2=\Om^2\l \bar g_{\mu \nu}dx^{\mu }dx^{\nu }-
(dz+kA^{\mu}dx_{\mu})^2\r ~.
\la{metgauge}
\eeq
We omit the graviscalar field in (\ref{tensors}) since it is not relevant
for our discussion. See \cite{KKdec1} for a discussion involving this
field.
Eq. (\ref{metgauge}) acquires the `usual' form for $A_{\mu}=0$.
For $A_{\mu }{\neq}0$, it is
invariant under the following transformations:

$$
x_{\mu }'=x_{\mu }~,~~~~~~~~z'=z+\ep (x_{\mu })~,
$$
\beq
A_{\mu }'=A_{\mu}-\fr{1}{k}\pl_{\mu }\ep ~.
\la{gaugeinv}
\eeq
Note that this is a $U(1)$ transformation for $A_{\mu }$, where $z$
plays the role of Goldstone field.
The $Z_2$ symmetry breaking creates the brane
and translational invariance in the fifth direction is spontaneously broken.
The breaking of the corresponding generator gives rise to a massive
$A_{\mu }$ field. By considering $z$ as $x_{\mu }$ dependent (which
corresponds to brane vibrations), the term
$(\pl_{\mu }z)^2$ (see below) appear in the $4$D action. This tell us that from the point
of view of $4$D observer, $z(x_{\mu })$ is a Goldstone field
which becomes the longitudinal component of $A_{\mu }$.
{}From this discussion it is clear that the fields $z(x_{\mu })$ and
$A_{\mu }$ reside on the $4$D brane.

We now calculate
the graviphoton mass. Taking into account (\ref{conf2}),
for the Einstein Tensor

\beq
{\cal G}_{MN}=R_{MN}-\fr{1}{2}G_{MN}R~,
\la{einsten}
\eeq
we have

$$
{\cal G}^G_{MN}={\cal G}^g_{MN}+(D-2)\l \na_M\ln \Om \na_N \ln \Om-
\na_M \na_N \ln \Om\r +
$$
\beq
(D-2)g_{MN}\l \na_P \na^P \ln \Om+\fr{1}{2}(D-3)\na_P\ln \Om \na^P\ln
\Om\r~,
\la{transf}
\eeq
where ${\cal G}^G$ and ${\cal G}^g$ are calculated using $G$ and $g$
respectively. The covariant derivatives $\na_M$ are built
from $g$, such that for a scalar function ${\cal S}$  

\beq
\na_M {\cal S}=\pl_M {\cal S}~,
\la{covdr1}
\eeq
while for a vector ${\cal V}$ 

\beq
\na_M {\cal V}^{\hs{0.4mm}N}=\pl_M {\cal V}^{\hs{0.4mm}N}+
\Ga^N_{MP}{\cal V}^{\hs{0.4mm}P}~,~~~
\na_M {\cal V}_N=\pl_M {\cal V}_N-\Ga^P_{MN}{\cal V}_P~.
\la{covdr2}
\eeq
{}From (\ref{einsten}) we have
\beq
R=-\fr{2}{D-2}G^{MN}{\cal G}_{MN}~,
\la{rel}
\eeq
and taking into account (\ref{transf}), we get:

$$
R(G)=\Om^{-2}\l R(g)-2(D-1)\na_M\na^M\ln \Om -
(D^2-3D+2)\na_M\ln \Om \na^M\ln \Om \r=~,
$$
\beq
~~\Om^{-2}\l R(g)-2(D-1)\fr{\na_M\na^M\Om }{\Om } -
(D^2-5D+4)\fr{\na_M \Om \na^M \Om }{\Om^2} \r~.
\la{curv}
\eeq

Calculating $R(g)$ through (\ref{tensors}) and keeping only relevant terms, we
have

$$
\sq{G}=\sq{-\bar g}\hs{1mm}\Om^5~,~~~~
$$
\beq
R(g)=\bar R(\bar g)
%-\fr{2}{3}\pl_{\mu}\ln \phi \hs{0.5mm}\pl^{\mu }\ln \phi 
+\fr{k^2}{4} F_{\mu \nu}F^{\mu \nu}+\dots~,~~~~
\la{decrel}
\eeq
where

\beq
F_{\mu \nu}=\pl_{\mu }A_{\nu}-\pl_{\nu}A_{\mu}~,
\la{strength}
\eeq
and $\bar R(\bar g)$ is the $4$D curvature, built from the physical $4$D metric
$\bar g_{\mu \nu}$.

Taking into account (\ref{covdr1}), (\ref{covdr2}) and (\ref{tensors}), we
have:

\beq
\na_M\na^M\Om =\l \pl^{\mu }\pl_{\mu }\Om +
2kA^{\mu }\pl_{\mu}\Om \hs{0.4mm}'+(k^2A^{\mu }A_{\mu }-
1)\Om \hs{0.4mm}''+\dots \r~,
\la{decder1}
\eeq

\beq
\na_M\Om \na^M\Om =\l \pl^{\mu}\Om \pl_{\mu}\Om+
2k\Om \hs{0.4mm}'A^{\mu }\pl_{\mu }\Om +(k^2A^{\mu }A_{\mu}-
1 )(\Om \hs{0.4mm}')^2\r~,
\la{decder2}
\eeq
where primes here denote derivatives with respect to $z$.
Using

\beq
\pl_{\mu }=\fr{\pl \bar z}{\pl x^{\mu }}\fr{\pl }{\pl z}=
\pl_{\mu}\bar z\cdot \fr{\pl }{\pl z}~,~~~~~~~~
\bar z\equiv z(x_{\mu })~,
\la{golds}
\eeq
{}from (\ref{curv}), (\ref{decder1}) and (\ref{decder2}) it finally 
follows that

$$
R(G)=\Om^{-2}R(g)-\Om^{-2}\l 2(D-1)\fr{\Om \hs{0.4mm}''}{\Om }+
(D^2-5D+4)\fr{\Om \hs{0.4mm}'\hs{0.3mm}^2}{\Om^2}\r \times
$$
\beq
\l\pl^{\mu }\bar z\pl_{\mu }\bar z+2kA^{\mu }\pl_{\mu }\bar z+
k^2A^{\mu }A_{\mu }-1 \r~.
\la{curvgAz}
\eeq
{}From (\ref{curvgAz}) we see that the field $\bar z$ can be absorbed
by $A_{\mu }$ by a suitable $U(1)$ transformation.

{}From the Einstein equation (\ref{eins}) we have:

\beq
-(\La+V)=-\fr{M^3}{2}\fr{D-2}{D}R+\fr{1}{2}\fr{D-2}{D}G^{AB}
(\pl_A\Phi^*\pl_B\Phi+\pl_B\Phi^*\pl_A\Phi)
\la{potconst}
\eeq
and substituting this in (\ref{act}), we get: 

\beq
S=\int d^5x\sq{G}\left [-\fr{M^3}{2}\fr{2(D-1)}{D}R+
\fr{2(D-1)}{D} \fr{1}{2}G^{AB}(\pl_A\Phi^*\pl_B\Phi+
\pl_B\Phi^*\pl_A\Phi) \right ]~.
\la{substact}
\eeq
With

\beq
\fr{1}{2}G^{AB}(\pl_A\Phi^*\pl_B\Phi+\pl_B\Phi^*\pl_A\Phi)=
\Om^{-2}v^2\te \hs{0.4mm}'\hs{0.3mm}^2
\l\pl^{\mu }\bar z\pl_{\mu }\bar z+2kA^{\mu }\pl_{\mu }\bar z+
k^2A^{\mu }A_{\mu }-1 \r~,
\la{deckin}
\eeq
After integrating over the fifth
dimension in (\ref{substact}), we obtain the reduced $4$D action:

\beq
S^{(4)}=\int d^4x\sq{-\bar g}\l -\fr{M_{\rm Pl}^2}{2}\bar R(\bar g)-T-
\fr{k^2}{4}B_{\mu \nu}B^{\mu \nu}+M_V^2(B_{\mu }+
\fr{1}{k}\pl_{\mu }{\cal Z})(B^{\mu }+
\fr{1}{k}\pl^{\mu }{\cal Z})\r ~,
\la{4Daction}
\eeq
where the $4$D Planck mass is

\beq
M_{\rm Pl}^2=\fr{2(D-1)}{D}M^3\int \Om^3dz ~,~~
\la{redpl}
\eeq
the $4$D brane tension is

\beq
T=M^3\fr{D-1}{D}\int \Om^3\left[2(D-1)\fr{\Om ''}{\Om }+
(D^2-5D+4)\fr{\Om '^2}{\Om^2}+\fr{2v^2}{M^3}\te '^2 \right]dz
\la{redT}
\eeq
and the mass of the graviphoton is

\beq
M_V^2=\fr{M^3}{M_{\rm Pl}^2}\fr{D-1}{D}k^2
\int  \Om^3\left[2(D-1)\fr{\Om''}{\Om }+
(D^2-5D+4)\fr{\Om '^2}{\Om^2}+\fr{2v^2}{M^3}\te '^2 \right]dz ~.
\la{massV}
\eeq
In obtaining (\ref{4Daction}) we have used

\beq
%\lan \phi \ran =1~,~~~~
B_{\mu }=M_{\rm Pl}A_{\mu }~,~~~~
B_{\mu \nu }=M_{\rm Pl}F_{\mu \nu}~,~~~~{\cal Z}=M_{\rm Pl}\bar z ~.
\la{redfields}
\eeq
Comparing (\ref{redT}) and (\ref{massV}),

\beq
M_V^2=\fr{T}{M_{\rm Pl}^2}k^2=\fr{T}{g_V^2M_{\rm Pl}^2} ~.
\la{mastens}
\eeq
%(\ref{mastens}) resembles famous relation for D-branes, in week coupling
%limit (???).

Simplifying (\ref{redT}) yields:

$$
T=\fr{4}{5}M^3\int dyA^2\l 4\fr{A_y''}{A}+\fr{A_y'^2}{A^2}+
\fr{2v^2}{M^3}\te_y'^2\r =
$$
\beq
\fr{16}{5}\sq{2\lam }M^3A_0^2m\l 2\al^2 I(2\al )+
(\al /2 -2\al^2)I(2\al +2) \r ~,
\la{tens1}
\eeq
where we have put $D=5$, the subscript $y$ denotes derivatives with respect to $y$,
and

\beq
I(\al )=\int_{0}^{1} \l \fr{1-\rho^2}{1+\rho^2} \r^{\al }
\fr{d\rho }{1-\rho^2}=\int_{0}^{\fr{\pi }{4}} \l \cos 2t\r^{\al -1}dt~.
\la{int}
\eeq
is some finite number whose value depends on the positive parameter $\al $. For   
$\al=1,~ I=\pi /4$, and for $\al=2,~ I=1/2$.

Note that the relation (\ref{mastens}) between the graviphoton mass and brane
tension, has same form as for
models with non warped geometry \cite{goldth}.

Simplifying (\ref{redpl}) one finds:

\beq
M_{\rm Pl}^2=\fr{8}{5}M^3\int_{-\inf }^{+\inf} A(y)dy=M^3R_{eff}~,
\la{4pl}
\eeq
where

\beq
R_{eff}=\fr{8}{5\sq{2\lam }}\int_{-\inf }^{+\inf} A(\xi)d\xi =
\fr{8A_0}{5\sq{2\lam }}\int_{-\inf }^{+\inf}
[\cosh(m\xi+\delta)]^{-\al }d\xi =
\fr{32A_0}{5m\sq{2\lam }}~I(\al )~.
\la{effR}
\eeq
Thus, even though the extra dimension $y$ is non-compact, its `effective'
size $R_{eff}$ 
is finite. In this sense the extra space is effectively compact.
Expression (\ref{4pl}) resembles the well known relation
$M_{\rm Pl}^2\sim M^{2+n}L^n$ (for $n=1$), which relates the effective $4$D
Planck scale to the fundamental scale $M$ and the volume ($\sim L^n$)
of the $n$ extra dimensions \cite{tevgr1, tevgr2}.
The crucial difference from models \cite{tevgr1, tevgr2} is that even for
values 
$M\sim M_{\rm Pl},~ R_{eff}\sim 1/M_{\rm Pl}$ in (\ref{4pl}), the 
desired hierarchy is obtained, thanks to the warped geometry. 
%At the end let us note, that we expect that various phenomenological and
%astrophysical implications \cite{goldth, impbr}, studied recently for the
%brane
%world universe, will also hold for our considered model. 

In conclusion, it would be interesting to investigate the possibility
of introducing a second domain wall, located at a suitable distance from
the first and characterized by the TeV scale . 'Double wall' solutions
that
are dynamically stabilized in axion type models with Minkowski 
background have been studied in ref. \cite{ourz2}. Some extension of the
model
considered here may well be required to implement such a scenario.

\bibliographystyle{unsrt}

\begin{thebibliography}{99}


%1
%theories with TeV scale Grav.

\bibitem{tevgr1}
N. Arkani-Hamed, S. Dimopoulos, G. Dvali,
Phys. Lett., B 429 (1998) 263;
I. Antoniadis, N. Arkani-Hamed, S. Dimopoulos, G. Dvali,
Phys. Lett., B 436 (1998) 257.

%2
% phenomenological impl. of TeV scale models.

\bibitem{tevgr2}
N. Arkani-Hamed, S. Dimopoulos, G. Dvali,
Phys. Rev., D 59 (1999) 086004;

%3 our Univ. as brane

\bibitem{univ}
K. Akama, Lect. Notes Phys. 176 (1982) 267; hep-th/0001113;
V. Rubakov, M. Shaposhnikov, Phys. Lett., B 125 (1983) 136;
A. Barnaveli, O. Kancheli, Sov. J. Nucl. Phys., 52 (1990) 576.


%4
% hierarchu with warped geometry

\bibitem{RS}   
L. Randall, R. Sundrum,
Phys. Rev. Lett., 83 (1999) 3370.

%5
% non-factorizable solution

\bibitem{gogb}   
M.Gogberashvili,
hep-ph/9812296; Europhys. Lett. 49 (2000) 396.

%6
% open compact dimensions

\bibitem{open1}   
L. Randall, R. Sundrum,
Phys. Rev. Lett., 83 (1999) 4690.

%7
% open dimensions for many branes

\bibitem{open2}   
N. Arkani-Hamed, S. Dimopoulos, G. Dvali, N. Kaloper,
Phys. Rev. Lett., 84 (2000) 586.

%8
% brane from sin-gordon
   
\bibitem{singor} 
A. Davidson, P. Mannheim, hep-th/0009064.


\bibitem{realscals}
W. Goldberger, H. Wise, hep-ph/9907218, hep-ph/9907447;
S. Ichinose, hep-th/0003275;
T. Gherghetta, E. Roessl, M. Shaposhnikov, hep-th/0006251;
A. Kehagias, K. Tamvakis, hep-th/0010112, hep-th/0011006;  
A. Iglesias, Z. Kakushadze, hep-th/0011111.


%9
% counting parameters

\bibitem{pars}
O. DeWolfe, D. Freedman, S. Gubser, A. Karch, hep-th/9909134.

%10
% model with Z_2

\bibitem{z2}
J. Preskill, S. Trevedi, F. Wilczeck, M. Wise,
Nucl. Phys., B 363 (1991) 207.

%11
% our paper - Z_2
  
\bibitem{ourz2}
G. Dvali, J. Nanobashvili, Z. Tavartkiladze,
Phys. Lett., B 352 (1995) 214.

%12

\bibitem{kib}
T.W.B. Kibble, G. Lazarides and Q. Shafi, Phys. Rev., D 26 (1982) 435; 
A. Vilenkin and A.E. Everett, Phys. Rev. Lett., 48 (1982) 1867;
G. Lazarides and Q. Shafi, Phys. Lett., B 115 (1982) 21. 


%13
% KK decomp. in RS-type models

\bibitem{KKdec1}
G. Kang, Y. Myung, hep-th/0007197.

%14

\bibitem{KKdec2}
M. Cvetic, H. Lu, C. Pope, hep-th/0009183;
Z. Kakushadze, P. Langfelder, hep-th/0011245.



%15
% goldstone theorem for extra space & graviphoton mass

\bibitem{goldth}   
G. Dvali, M. Shifman,
Phys. Rept., 320 (1999) 107; hep-th/9904021;
M. Bando at el., hep-ph/9906549;
G. Dvali, I. Kogan, M. Shifman, hep-th/0006213;
A. Dobado, A. Maroto, hep-ph/0007100.

%16
%various  phen. & astro. implications for brane universe models
%
%\bibitem{impbr}
%G. Dvali, M. Shifman, Phys. Lett., B 475 (2000) 295.



\end{thebibliography}

\end{document}